\documentclass[12pt]{article}
\usepackage{amsbsy}
\usepackage{amsfonts}
\usepackage{amsmath}
\usepackage{amssymb}
\usepackage{apacite}
\usepackage{setspace}
\usepackage{graphicx}
\usepackage{bm,multicol}
\usepackage[english]{babel}
\usepackage{natbib}
\usepackage[T1]{fontenc}
\usepackage[utf8]{inputenc}
\usepackage{authblk}
\usepackage{xcolor}
\usepackage{hyperref}
\newcommand{\newc}{\newcommand}
\newc{\N}{\mbox{N}}
\newc{\1}{\bf{1}}

\begin{document}
\title{Bayesian testing of linear versus nonlinear effects using Gaussian process priors}

\author{Joris Mulder}
%\author{}

\date{}

\thispagestyle{empty} \maketitle

\begin{abstract}
A Bayes factor is proposed for testing whether the effect of a key predictor variable on the dependent variable is linear or nonlinear, possibly while controlling for certain covariates. The test can be used (i) when one is interested in quantifying the relative evidence in the data of a linear versus a nonlinear relationship and (ii) to quantify the evidence in the data in favor of a linear relationship (useful when building linear models based on transformed variables). Under the nonlinear model, a Gaussian process prior is employed using a parameterization similar to Zellner's $g$ prior resulting in a scale-invariant test. Moreover a Bayes factor is proposed for one-sided testing of whether the nonlinear effect is consistently positive, consistently negative, or neither. Applications are provides from various fields including social network research and education.
\end{abstract}

\section{Introduction}
Linearity between explanatory and dependent variables is a key assumption in most statistical models. In linear regression models, the explanatory variables are assumed to affect the dependent variables in a linear manner, in logistic regression models it is assumed that the explanatory variables have a linear effect on the logit of the probability of a success on the outcome variable, in survival or event history analysis a log linear effect is generally assumed between the explanatory variables and the event rate, etc. Sometimes nonlinear functions (e.g., polynomials) are included of certain explanatory variables (e.g., for modeling curvilinear effects), or interaction effects are included between explanatory variables, which, in turn, are assumed to affect the dependent variable(s) in a linear manner.

Despite the central role of linear effects in statistical models, statistical tests of linearity versus nonlinearity are only limitedly available. In practice researchers tend to eyeball the relationship between the variables based on a scatter plot. When a possible nonlinear relationship is observed, various linear transformations (e.g., polynomial, logarithmic, Box-Cox) are applied and significance tests are executed to see if the coefficients of the transformed variables are significant or not. Eventually, when the nonlinear trend results in a reasonable fit, standard statistical inferential methods are applied (such as testing whether certain effects are zero and/or evaluating interval estimates).

%the focus is typically on testing whether some nonlinear function of an explanatory variable $x$ is different from zero, e.g., when testing $H_0:\beta_{x^2}=0$ versus $H_1:\beta_{x^2}\not=0$, where $\beta_{x^2}$ denotes the coefficient of the $x$ squared. When the null hypothesis can be rejected, the conclusion is that there is evidence that the effect of $x$ is nonlinear (in this case curvilinear with a second order polynomial). Then, typically based on eyeballing, one determines whether the polynomial seems to reasonably fit the nonlinear relationship in the data, and, if not, additional nonlinear terms are included, and additional statistical tests are executed to determine whether other terms are significantly different from 0.

This procedure is problematic for several reasons. First, executing many different significance tests on different transformed variables may result in $p$-hacking and inflated type I errors. Second, regardless of the outcome of a significance test, e.g., when testing whether the coefficient of the square of the predictor variable, $X^2$, equals zero, $H_0:\beta_{X^2}=0$ versus $H_1:\beta_{X^2}\not=0$, we would not learn whether $X^2$ has a linear effect on $Y$ or not; only whether an increase of $X^2$ results \textit{on average} in an increase/decrease of $Y$ or not. Third, nonlinear transformations (e.g., polynomials, logarithmic, Box-Cox) are only able to create approximate linearity for a limited set of nonlinear relationships. Fourth, eyeballing the relationship can be subjective, and instead a principle approach is needed.

To address these shortcomings this paper proposes a Bayes factor for the following hypothesis test
\begin{eqnarray}
\nonumber\text{$M_0:$ ``$X$ has a linear effect on $Y$''}~~\\
\label{Htest}\text{versus}~~~~~~~~~~~~~~~~~~~\\
\nonumber \text{$M_1:$ ``$X$ has a nonlinear effect on $Y$'',}
\end{eqnarray}
possibly while controlling for covariates. Unlike $p$ value significance tests, a Bayes factor can be used for quantifying the relative evidence in favor of linearity \citep{Wagenmakers:2007}. Furthermore, Bayes factors are preferred for large samples as significance tests may indicate that the null model needs to be rejected even though inspection may not show striking discrepancies from linearity. This behavior is avoided when using Bayesian model selection \citep{Raftery:1995}.

Under the alternative model, a Gaussian process prior is used to model the nonlinear effect. A Gaussian process is employed due to its flexibility to model nonlinear relationships \citep{Rasmussen:2007}. Because nonlinear relationships are generally fairly smooth, the Gaussian process is modeled using a squared exponential kernel. Furthermore, under both models a $g$ prior approach is considered \cite{Zellner:1986} so that the test is scale-invariant of the dependent variable. To our knowledge a $g$ prior was not used before for parameterizing a Gaussian process. As a result of the common parameterization under both models, the test comes down to testing whether a specific scale parameter equals zero or not, where a zero value implies linearity. Under the alternative the scale parameter is modeled using a half-Cauchy prior with a scale hyperparameter that can be chosen depending on the expected deviation from linearity under the alternative model. %The hyperparameter is chosen such that it corrects for the scale of the predictor variable.

Furthermore, in the case of a nonlinear effect, a Bayes factor is proposed for testing whether the effect is consistently increasing, consistently decreasing or neither. This test can be seen as a novel nonlinear extension to one-sided testing. 

Finally note that the literature on Gaussian processes has mainly focused on estimating nonlinear effects \citep[e.g.,][]{Rasmussen:2007,Duvenaud:2011,Cheng:2019}, and not testing nonlinear effects, with an exception of \cite{Liu:2017} who proposed a significance (score) test, which has certain drawbacks as mentioned above. Further note that spline regression analysis is also typically used for estimating nonlinear effects, and not for testing (non)linearity.

The paper is organized as follows. Section 2 describes the linear and nonlinear Bayesian models and the corresponding Bayes factor. Its behavior is also explored in a numerical simulation. Section 3 describes the nonlinear one-sided Bayesian test. Subsequently, Section 4 presents 4 applications of the proposed methodology in different research fields. We end the paper with a short discussion in Section 5.

\section{A Bayes factor for testing (non)linearity}
\subsection{Model specification}
Under the standard linear regression model, denoted by $M_0$, we assume that the mean of the dependent variable $Y$ depends proportionally on the key predictor variable $X$, possibly while correcting for certain covariates. Mathematically, this implies that the predictor variable is multiplied with the same coefficient, denoted by $\beta$, to compute the (corrected) mean of the dependent variable for all values of $X$. The linear model can then be written as
\begin{equation}
M_0:\textbf{y}\sim\mathcal{N}(\beta\textbf{x} + \bm\gamma\textbf{Z},\sigma^2 \textbf{I}_n),
\end{equation}
where $\textbf{y}$ is a vector containing the $n$ observations of the dependent variable, $\textbf{x}$ contains the $n$ observations of the predictor variable, $\textbf{Z}$ is a $n\times k$ matrix of covariates (which are assumed to be orthogonal to the key predictor variable) with corresponding coefficients $\bm\gamma$, and $\sigma^2$ denotes the error variance which is multiplied with the identity matrix of size $n$, denoted by $\textbf{I}_n$. To complete the Bayesian model, we adopt the standard $g$ prior approach \citep{Zellner:1986} by setting a Gaussian prior on $\beta$ where the variance is scaled based on the error variance, the scale of the predictor variable, and the sample size, with a flat prior for the nuisance regression coefficients, and the independence Jeffreys prior for the error variance, i.e.,
\begin{eqnarray*}
\beta |\sigma^2 &\sim & N(0,\sigma^2g(\textbf{x}'\textbf{x})^{-1})\\
p(\bm\gamma) & \propto & 1\\
p(\sigma^2) & \propto & \sigma^{-2}.
\end{eqnarray*}
The prior mean is set to the default value of 0 so that, a priori, small effects in absolute value are more likely than large effects (as is common in applied research) and positive effects are equally likely as negative effects (an objective choice in Bayesian one-sided testing \citep{Jeffreys,Mulder:2010}). By setting $g=n$ we obtain a unit-information prior \citep{KassWasserman:1995,Liang:2008} which will be adopted throughout this paper\footnote{Note that we don't place a prior on $g$, as is becoming increasingly common \citep{Liang:2008,Rouder:2009,Bayarri:2007}, because we are not specifically testing whether $\beta$ equals 0 and to keep the model as simple as possible.}.
%[[[Mention orthogonality of Z t.o.v. x? We could make the nuisance parameter orthogonal to x by setting $Z = (I - \textbf{x}(\textbf{x}'\textbf{x})^{-1}\textbf{x}')$ \citep[e.g., see][]{Bayarri:2007}]]]}.

Under the alternative nonlinear model, denoted by $M_1$, we assume that the mean of the dependent variable does not depend proportionally on the predictor variable. This implies that the observations of the predictor variable can be multiplied with different values for different values of the predictor variable $X$. This can be written as follows
\begin{equation}
M_1:\textbf{y}\sim\mathcal{N}(\bm\beta(\textbf{x})\circ\textbf{x} + \bm\gamma\textbf{Z},\sigma^2 \textbf{I}_n),
\end{equation}
where $\bm\beta(\textbf{x})$ denotes a vector of length $n$ containing the coefficients of the corresponding $n$ observations of the predictor variable $\textbf{x}$, and $\circ$ denotes the Hadamard product. The vector $\bm\beta(\textbf{x})$ can be viewed as the $n$ realizations when plugging the different values of $\textbf{x}$ in a unknown theoretical function $\beta(x)$. Thus, in the special case where $\beta(x)$ is a constant function, say, $\beta(x)=\beta$, model $M_1$ would be equivalent to the linear model $M_0$.

Next we specify a prior probability distribution for the function of the coefficients. Because we are testing for linearity, it may be more likely to expect relatively smooth changes between different values, say, $\beta_i(x_i)$ and $\beta_j(x_j)$ than large changes when the values $x_i$ and $x_j$ are close to each other. A Gaussian process prior for the function $\bm\beta(\textbf{x})$ has this property which is defined by % changes of  function values are changes to (far away from) each other when the plugged in $X$ values are close to (far away from) each other. This would be equivalent with assuming that relatively small violations of linearity are more likely a priori than large violations. Note that this corresponds to one of the desiderata of \cite{Jeffreys} for prior specification. In the current setting when testing for linearity versus nonlinearity a Gaussian process prior abides this property, and is defined by
\begin{equation}
\bm\beta(\textbf{x}) | \tau^2,\xi \sim \mathcal{GP}(\textbf{0},\tau^2k(\textbf{x},\textbf{x}'|\xi)),
\end{equation}
which has a zero mean function and a kernel function $k(\cdot,\cdot)$ which defines the covariance of the coefficients as a function of the distance between values of the predictor variable. A squared exponential kernel will be used which is given by
\begin{equation}
\label{kernel}
k(x_i,x_j|\xi) = \exp\left\{ -\tfrac{1}{2}\xi^2(x_i-x_j)^2 \right\},
\end{equation}
for $i,j=1,\ldots,n$. As can be seen, predictor variables $x_i$ and $x_j$ that are close to (far away from) each other have a larger (smaller) covariance, and thus, are on average closer to (further away from) each other. The hyperparameter $\xi$ controls the smoothness of the function where values close to 0 imply very smooth function shapes and large values imply highly irregular shapes (as will be illustrated later). Note that typically the smoothness is parameterized via the reciprocal of $\xi$. Here we use the current parameterization so that the special value $\xi=0$ would come down to a constant function, say $\beta(x)=\beta$, which would correspond to a linear relationship between the predictor and the outcome variable.

The hyperparameter $\tau^2$ controls the prior magnitude of the coefficients, i.e., the overall prior variance for the coefficients. We extend the $g$ prior formulation to the alternative model by setting $\tau^2=\sigma^2g(\textbf{x}'\textbf{x})^{-1}$ and specify the same priors for $\bm\gamma$ and $\sigma^2$ as under $M_0$. Furthermore, by taking into account that the Gaussian process prior implies that the coefficients for the observed predictor variables follow a multivariate normal distribution, the priors under $M_1$ given the predictor variables can be formulated as 
\begin{eqnarray*}
\bm\beta(\textbf{x})|\sigma^2,\xi,\textbf{x} & \sim & \mathcal{N}(\textbf{0}, \sigma^2g(\textbf{x}'\textbf{x})^{-1}k(\textbf{x},\textbf{x}'|\xi))\\
p(\bm\gamma) & \propto & 1\\
p(\sigma^2) & \propto & \sigma^{-2}.
\end{eqnarray*}
To complete the model a half-Cauchy prior is specified for the key $\xi$ having prior scale $s_{\xi}$, i.e.,
\[
\xi \sim  \text{half-}\mathcal{C}(s_{\xi}).
\]
The motivation for this prior is based on one of \cite{Jeffreys} desiderata which states that small deviations from the null value are generally more likely a priori than large deviations otherwise there would be no point in testing the null value. In the current setting this would imply that small deviations from linearity are more likely to be expected than large deviations. This would imply that values of $\xi$ close to 0 are more likely a priori than large values, and thus that the prior distribution for $\xi$ should be a decreasing function. The half-Cauchy distribution satisfies this property. Further note that the half-Cauchy prior is becoming increasingly popular for scale parameters in Bayesian analyses \citep{Gelman:2006,Polson:2012,MulderPericchi:2018}.

\begin{figure}[t!]
\centering\includegraphics[height=.8\textwidth]{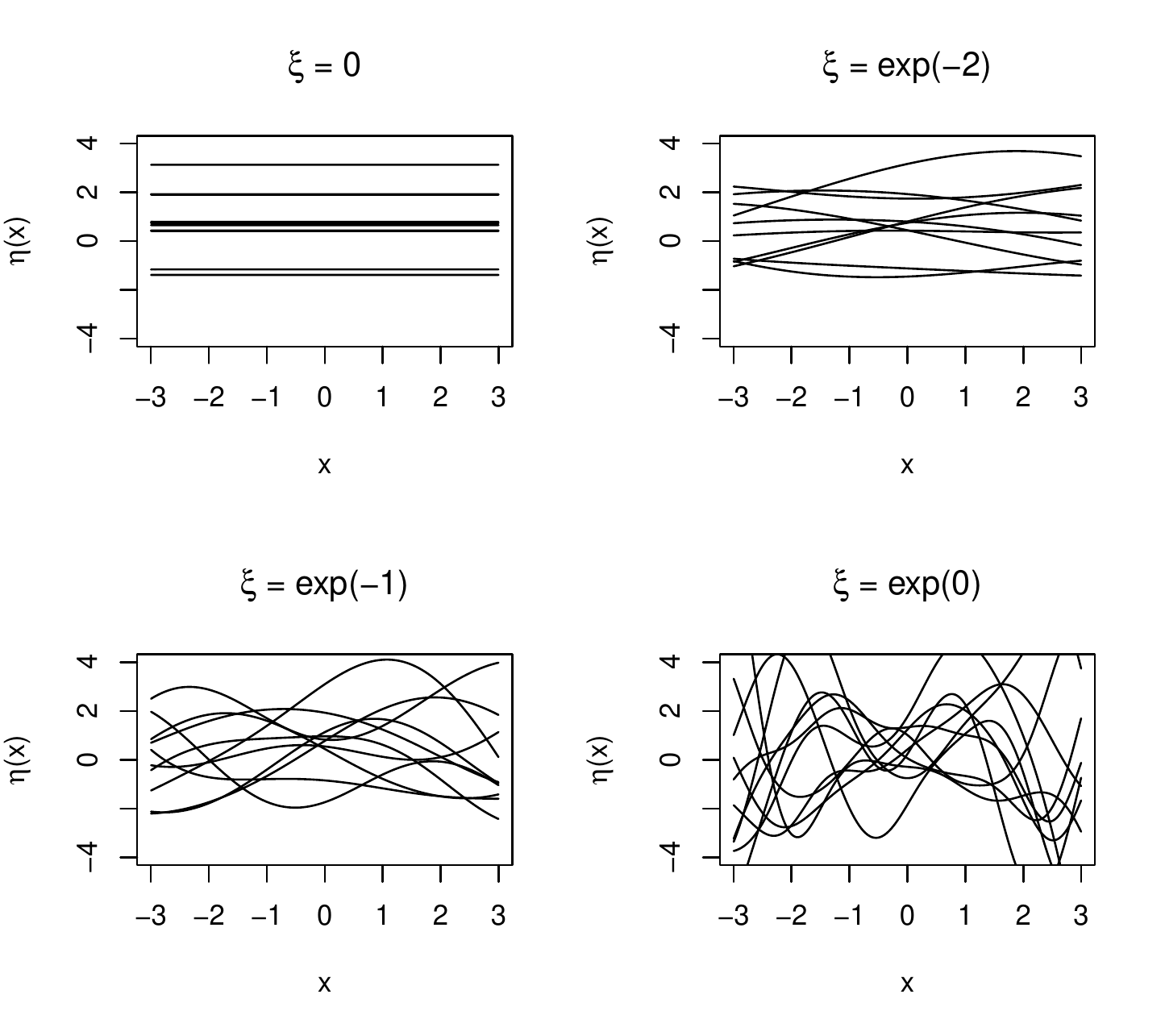}
\caption{Ten random slope functions $\bm\eta(\textbf{x})$ when using a Gaussian process prior on $\bm\beta(\textbf{x})$ with mean $\textbf{0}$ and squared exponential kernel with $\tau^2=1$ and $\xi=0$ (constant effect), $\xi=\exp(-2)$ (small deviations from a constant effect), $\xi=\exp(0)=1$ (medium deviations from a constant effect), and $\xi=\exp(0)$ (large deviations from a constant effect).}
\label{fig1}
\end{figure}

The prior scale for key parameter $\xi$ under $M_1$ should be carefully specified as it defines which deviations from linearity are most plausible. To give the reader more insight about how $\xi$ affects the distribution of the slopes of $\textbf{y}$ as function of $\textbf{x}$, Figure \ref{fig1} displays 10 random draws of the function of slopes when setting $\xi_1=0$ (Figure \ref{fig1}a), $\xi=\exp(-2)$ (Figure \ref{fig1}b), $\xi=\exp(-1)=1$ (Figure \ref{fig1}c), $\xi=\exp(0)$ (Figure \ref{fig1}d) while fixing $\tau^2=\sigma^2 g (\textbf{x}'\textbf{x})^{-1}=1$, where the slope function is defined by
\begin{equation}
\bm\eta (\textbf{x}) = \frac{d}{d\textbf{x}}[\bm\beta(\textbf{x})\circ\textbf{x}] = \bm\beta(\textbf{x}) + 
\frac{d}{d\textbf{x}}[\bm\beta(\textbf{x})]\circ\textbf{x}.
\end{equation}
The figure shows that by increasing $\xi$ we get larger deviations from a constant slope. Based on these plots we qualify the choices $\xi=\exp(-2)$, $\exp(-1)$, and 1 as small deviations, medium deviations, and large deviations from linearity, respectively.

Because the median of a half-Cauchy distribution is equal to the scale parameter $s_{\xi}$, the scale parameter could be set based on the expected deviation from linearity. It is important to note here that the expected deviation depends on the range of the predictor variable: In a very small range it may be expected that the effect is close to linear but in a wide range of the predictor variable, large deviations from linearity may be expected. Given the plots in Figure \ref{fig1}, one could set the prior scale equal to $s_{\xi} = 6e/\text{range}(\textbf{x})$, where $e$ can be interpreted as a standardized measure for the deviation from linearity such that setting $e = \exp(-2), \exp(-1)$, or $\exp(0)$ would imply small, medium, or large deviations from linearity, respectively. Thus, if the range of $\textbf{x}$ would be equal to 6 (as in the plots in Figure \ref{fig1}), the median of $\xi$ would be equal to $\exp(-2), \exp(-1)$, and $\exp(0)$, as plotted in Figure \ref{fig1}.

\subsection{Bayes factor computation}
The Bayes factor is defined as the ratio of the marginal (or integrated) likelihoods under the respective models. For this reason it is useful to integrate out the coefficient $\beta$ under $M_0$ and the coefficients $\bm\beta(\textbf{x})$ under $M_1$, which are in fact nuisance parameters in the test. This yields the following integrated models
\begin{align}
M_0 : &
\begin{cases} 
   \textbf{y} | \textbf{x},\bm\gamma,\sigma^2 \sim  \mathcal{N}(\textbf{Z}\bm\gamma,\sigma^2 g (\textbf{x}'\textbf{x})^{-1}\textbf{x}\textbf{x}'+\sigma^2\textbf{I}_n) \\
   p(\bm\gamma)  \propto  1\\
   p(\sigma^2) \propto \sigma^{-2}
  \end{cases}\\
M_1 : &
\begin{cases} 
   \textbf{y} | \textbf{x},\bm\gamma,\sigma^2,\xi \sim \mathcal{N}(\textbf{Z}\bm\gamma,\sigma^2 g (\textbf{x}'\textbf{x})^{-1} k(\textbf{x},\textbf{x}'|\xi) \circ \textbf{x}\textbf{x}'+\sigma^2\textbf{I}_n) \\
   p(\bm\gamma)  \propto  1\\
   p(\sigma^2) \propto \sigma^{-2}\\
   \xi \sim  \text{half-}\mathcal{C}(s_{\xi}),
  \end{cases}
\end{align}
As can be seen $\sigma^2$ is a common factor in all (co)variances of $\textbf{y}$ under both models. This makes inferences about $\xi$ invariant to the scale of the outcome variable. Finally note that the integrated models clearly show that the model selection problem can concisely be written as
\begin{eqnarray*}
M_0&:&\xi = 0\\
M_1&:&\xi > 0.
\end{eqnarray*}
because $k(\textbf{x},\textbf{x}'|\xi)=1$ when setting $\xi=0$.

Using the above integrated models, the Bayes factor can be written as
\begin{equation*}
B_{01} = \frac{
\iint p(\textbf{y}|\textbf{x},\bm\gamma,\sigma^2,\xi=0)\pi(\bm\gamma)\pi(\sigma^2)
d\bm\gamma d\sigma^2
}{
\iiint p(\textbf{y}|\textbf{x},\bm\gamma,\sigma^2,\xi)\pi(\bm\gamma)\pi(\sigma^2)\pi(\xi)d\bm\gamma d\sigma^2 d\xi
},
\end{equation*}
which quantifies the relative evidence in the data between the linear model $M_0$ and the nonlinear model $M_1$.
Different methods can be used for computing marginal likelihoods. Throughout this paper we use an importance sample estimate.  The R code for the computation of the marginal likelihoods and the sampler from the posterior predictive distribution can be found in the supplementary material.

\subsection{Numerical behavior}
\begin{figure}[t!]
\centering\includegraphics[height=.8\textwidth]{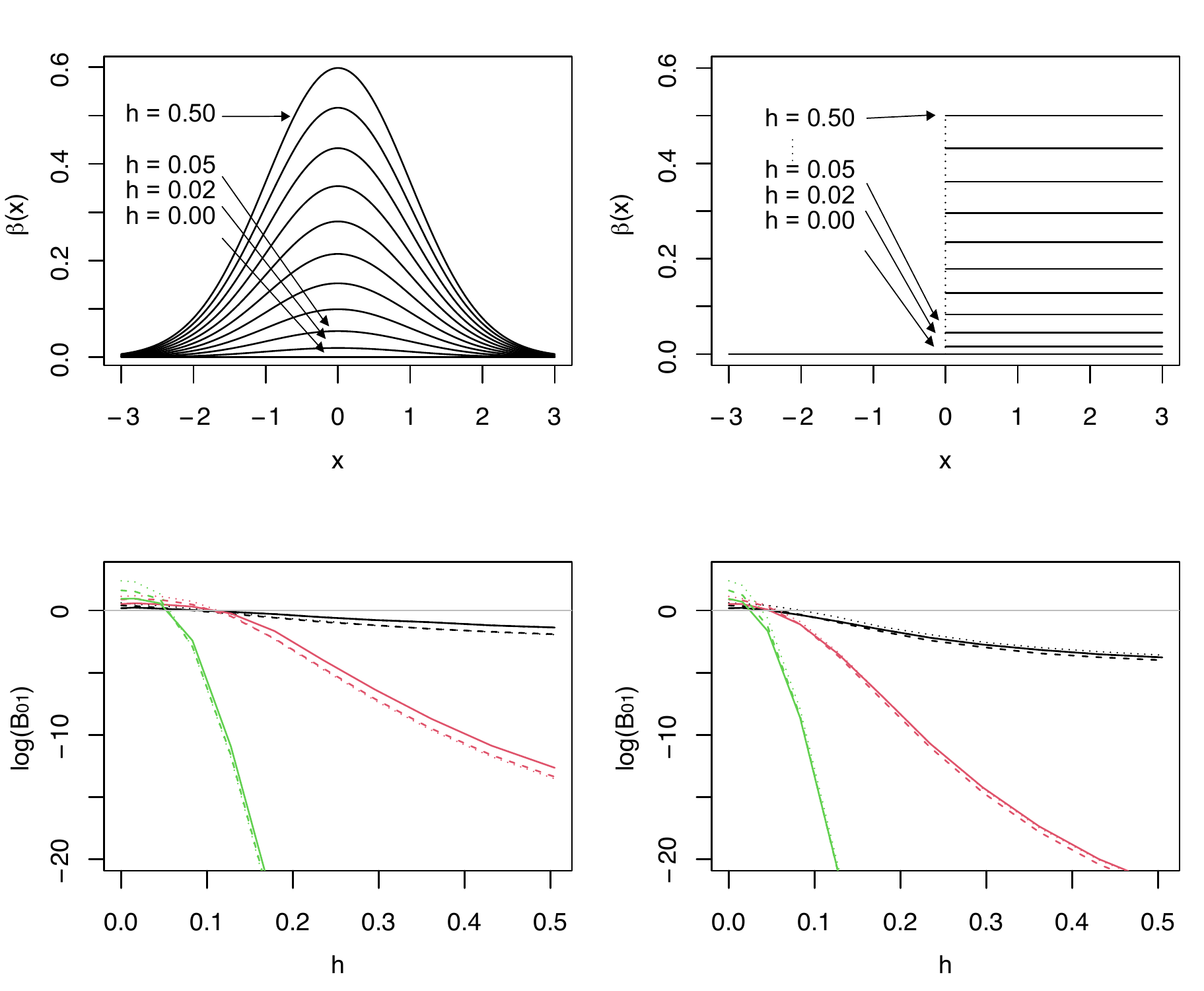}
\caption{Left panels. (Upper) Example functions of $\beta(x)=3h\phi(x)$, for $h=0,\ldots,.5$, where $\phi$ is the standard normal probability density function. Left lower panel and (lower) corresponding logarithm of the Bayes factor as function of $h$ for $n=20$ (black lines), $50$ (red lines), and 200 (green lines) for a small prior scale $s_{\xi}=\exp(-2)$ (solid lines), a medium prior scale $s_{\xi}=\exp(-1)$ (dashed lines), and a large prior scale $s_{\xi}=\exp(0)$ (dotted lines). Right panels. (Upper) Example functions of $\beta(x)=h~1(x>0)$, for $h=0,\ldots,.5$ and (lower) corresponding logarithm of the Bayes factor as function of $h$.}
\label{simulfig}
\end{figure}

Numerical simulation were performed to evaluate the performance of the proposed Bayes factor. The nonlinear function was set equal to $\beta(x)=3h\phi(x)$, for $h=0,\ldots,.5$, where $\phi$ is the standard normal probability density function (Figure \ref{simulfig}; upper left panel). In the case $h=0$, the effect is linear, and as $h$ increase, the effect becomes increasingly nonlinear. The dependent variable was computed as $\bm\beta(\textbf{x})\circ\textbf{x}+\bm\epsilon$, where $\bm\epsilon$ was sampled from a normal distribution with mean 0 and $\sigma=.1$.

The logarithm of the Bayes factor, denoted by $\log(B_{01})$, was computed between the linear model $M_0$ and the nonlinear model $M_1$ (Figure \ref{simulfig}; lower left panel) while setting the prior scale equal to $s_{\xi}=\exp(-2)$ (small prior scale; solid line), $\exp(-1)$ (medium prior scale; dashed line), and $\exp(0)$ (large prior scale; dotted line) for sample size $n=20$ (black lines), $50$ (red lines), and 200 (green lines) for equally distant predictor values in the interval $(-3,3)$. Overall we see the expected trend where we obtain evidence in favor of $M_0$ in the case $h$ is close to zero and evidence in favor of $M_1$ for larger values of $h$. Moreover the evidence for $M_0$ ($M_1$) is larger for larger sample sizes and larger prior scale when $h=0$ ($h\gg 0$) as anticipated given the consistent behavior of the Bayes factor.

Next we investigated the robustness of the test to nonlinear relationships that are not smooth as in the Gaussian processes having a squared exponential kernel. A similar analysis was performed when using the nonsmooth, discontinuous step function $\beta(x)=h~1(x>0)$, where $1(\cdot)$ is the indicator function, for $h=0,\ldots,.5$ (Figure \ref{simulfig}; upper right panel). Again the dependent variable was computed as $\bm\beta(\textbf{x})\circ\textbf{x}+\bm\epsilon$ and the logarithm of the Bayes factor was computed (Figure \ref{simulfig}; lower right panel). The Bayes factor shows a similar behavior as the above example where the data came from a smooth nonlinear alternative. The similarity of the results can be explained by the fact that even though the step function cannot be generated using a Gaussian process with a squared exponential kernel, the closest approximation of the step function is still nonlinear, and thus evidence is found against the linear model $M_0$ in the case $h>0$. This illustrates that the proposed Bayes factor is robust to nonsmooth nonlinear alternative models.

\section{Extension to one-sided testing}
When testing linear effects, the interest is often on whether the effect is either positive or negative if the null does not hold. Equivalently in the case of nonlinear effects the interest would be whether the effect is consistently increasing or consistently decreasing over the range of $X$. To model this we divide the parameter space under the nonlinear model $M_1$ in three subspaces:
\begin{eqnarray}
\nonumber M_{1,\text{positive}} &:& \text{``the nonlinear effect of $X$ on $Y$ is consistently positive''}\\
\nonumber M_{1,\text{negitive}} &:& \text{``the nonlinear effect of $X$ on $Y$ is consistently negative''}\\
\nonumber M_{1,\text{complement}}&:& \text{``the nonlinear effect of $X$ on $Y$ is neither consistently}\\
\label{onesided}&&\text{positive, nor consistently negative''.}
\end{eqnarray}
Note that the first model implies that the slope function is consistently positive, i.e., $\eta(x)>0$, the second implies that the slope is consistently negative, i.e., $\eta(x)<0$, while the third complement model assumes that the slope function is neither consistently positive nor negative.
%regarding the mean function of $Y$ is consistently positive over a certain region, the second model implies that the derivative is consistently negative, and the third model implies that the derivative is not consistently positive or negative. This derivative is given by
%\[
%\textbf{h}(\textbf{x})=\frac{d}{d\textbf{x}}[E\{\textbf{y}\}] = \frac{d}{d\textbf{x}}[\bm\beta(\textbf{x})\circ\textbf{x}] = \bm\beta(\textbf{x}) + 
%\frac{d}{d\textbf{x}}[\bm\beta(\textbf{x})]\circ\textbf{x}.
%\]
%Because the derivative of a Gaussian process is also a Gaussian process (Rasmussen \& Williams, 2006, Ch. 12????), the derivative of interest, $\textbf{h}(\textbf{x})$, is also a Gaussian process. Note that given a vector $\bm\beta(\textbf{x})$ of length $n$ this derivative is easy to obtain numerically.

Following standard Bayesian methodology using truncated priors for one-sided testing problems \citep{Klugkist:2005,Mulder:2020}, we set truncated Gaussian process priors on each of these three models, e.g., for model $M_{1,\text{positive}}$, this comes down to
\[
\pi_{1,\text{pos}}(\bm\beta(\textbf{x})|\xi_1,\tau_1) = \pi_1(\bm\beta(\textbf{x})|\xi,\tau)
\text{Pr}(\bm\eta(\textbf{x})>\textbf{0}|M_1,\xi,\tau)^{-1} 1_{\bm\eta(\textbf{x})>\textbf{0}}(\bm\beta(\textbf{x})),
\]
where $1_{\{\cdot\}}(\cdot)$ denotes the indicator function, and the prior probability, which serves as normalizing constant, equals
\[
\text{Pr}(\bm\eta(\textbf{x})>\textbf{0}|M_1,\xi,\tau) = \int_{\bm\eta(\textbf{x})>\textbf{0}} \pi_1(\bm\beta(\textbf{x})|\xi,\tau) d\bm\beta(\textbf{x}).
\]
Note that the prior probability for a consistently positive effect is equal because the prior mean of $\bm\beta(\textbf{x})$ equals $\textbf{0}$. Given this prior, the Bayes factor of each constrained model against the unconstrained model $M_1$ is then given by the ratio of the posterior and prior probabilities that the constraints hold under $M_1$, e.g.,
\[
B_{(1,\text{pos})u} = \frac{\text{Pr}(\bm\eta(\textbf{x})>\textbf{0}|M_1,\textbf{y})}{\text{Pr}(\bm\eta(\textbf{x})>\textbf{0}|M_1)}.
\]
Bayes factors between the above three models can then be computed using the transitive property of the Bayes factor, e.g., $B_{(1,\text{pos})(1,\text{comp})}=B_{(1,\text{pos})u}/B_{(1,\text{comp})u}$.

The choice of the prior of $\xi$ (which reflects the expected deviation from linearity before observing the data) implicitly determines the prior probability that the nonlinear effect is consistently positive or negative effect. This makes intuitive sense as large (small) deviations from linearity make it less (more) likely that the effect is either consistently positive or negative. This can also be observed from a careful inspection of the random draws in Figure \ref{fig1}. When $\xi=\exp(-2)$, we see that 4 out of 10 random functions in Figure \ref{fig1}b are consistently positive and 2 functions are consistently negative; when $\xi=\exp(-1)$ we see 1 random function that is consistently positive and 1 function that is consistently negative; and when $\xi=\exp(0)$ none of the 10 draws are either consistently positive or negative. The probabilities for a consistently positive (or negative) effect can simply be computed as the proportion of draws of random functions that is consistently positive (or negative). The choices $s_{\xi}=\exp(-2),~\exp(-1),$ and $\exp(0)$ result in prior probabilities for a consistently positive effect are approximately 0.25, 0.14, and 0.06.

\section{Empirical applications}
\subsection{Neuroscience: Facebook friends vs grey matter}
\cite{Kanai:2012} studied the relationship between the number of facebook friends and the grey matter density in regions of the brain that are related to social perception and associative memory to better understand the reason reasons for people to participate in online social networking. Here we analyze the data from the right entorhinal cortex ($n=41$). Due to the nature of the variables a positive relationship was expected. Based on a significance test \citep{Kanai:2012} and a Bayes factor \citep{Wetzels:2012} on a sample of size 41, it was concluded that there is evidence for a nonzero correlation between the square root of the number of Facebook friends and the grey matter density. In order for a correlation to be meaningful however it is important that the relationship is (approximately) linear. Here we test whether the relationship is linear or nonlinear. Furthermore, in the case of a nonlinear relationship, we test whether the relationships are consistently positive, consistently negative, or neither. Besides the predictor variable, the employed model has an intercept. The predictor variable is shifted to have a mean of 0 so that it is independent of the vector of ones for the intercept.

\begin{figure}[t!]
\centering\includegraphics[height=1\textwidth]{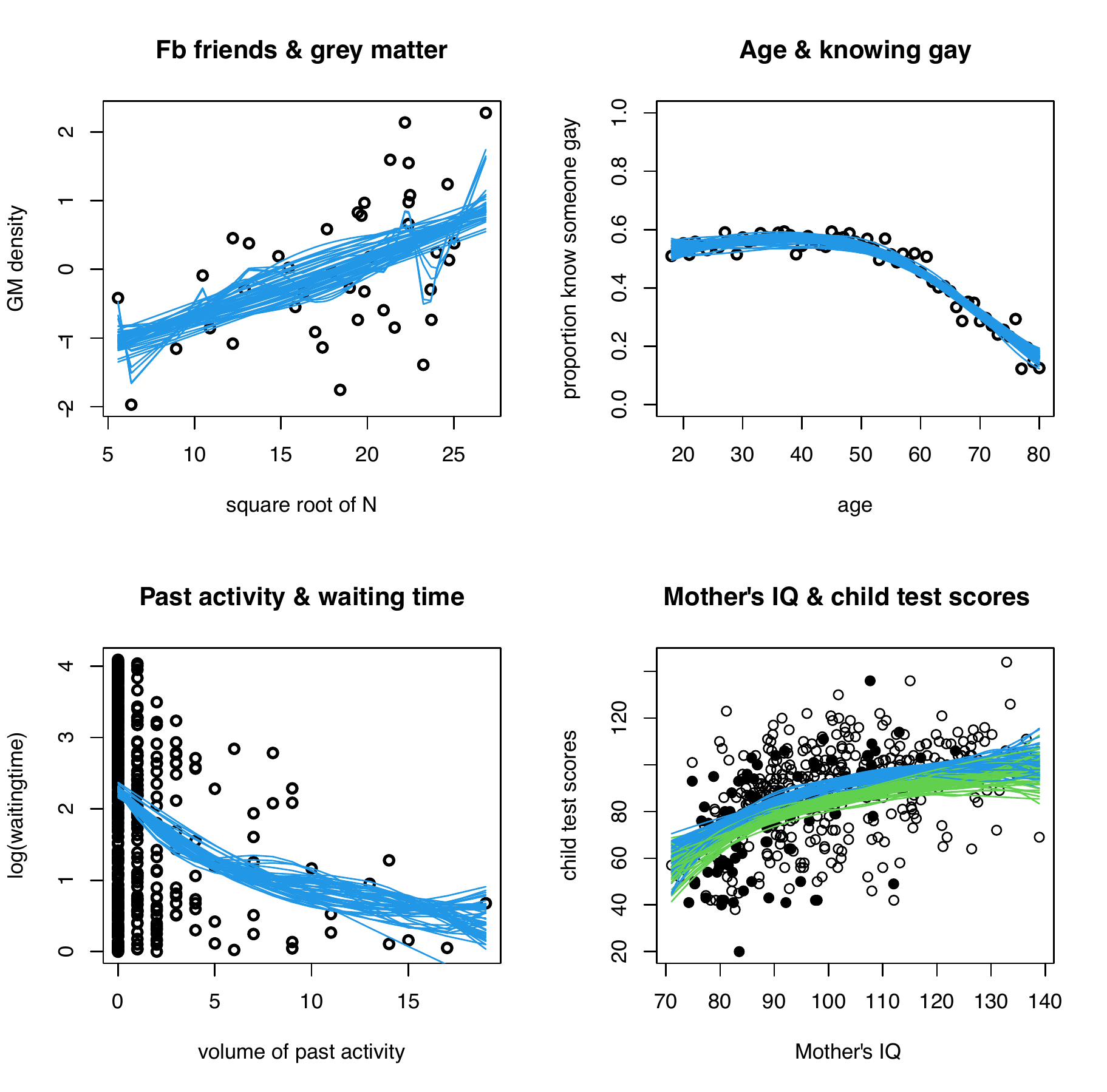}
\caption{Observations (circles) and 50 draws of the mean function (lines) under the nonlinear model $M_1$ for the four different applications. In the lower left panel draws are given in the case the mother finished her high school (light blue lines) or not (dark blue lines).}
\label{appfigure}
\end{figure}

The Bayes factor between the linear model against the nonlinear model when using a prior scale of $\exp(-1)$ (medium effect) was equal to $B_{01}=2.50$ (with $\log(B_{01})=0.917$). This implies very mild evidence for a linear relationship between the square root of the number of Facebook friends and grey matter density in this region of the predictor variable. When assuming equal prior model probabilities, this would result in posterior model probabilities of .714 and .286 for $M_0$ and $M_0$, respectively. Thus if we would conclude that the relation is linear there would be a conditional error probability of drawing the wrong conclusion. Table \ref{appresults} presents the Bayes factors also for the other prior scales which tell a similar tale. Figure \ref{appfigure} (upper left panel) displays the data (circles; replotted from Kanai et al., 2012) and 50 draws of the posterior distribution density for the mean function under the nonlinear model at the observed values of the predictor variable. As can be seen most draws are approximately linear, and because the Bayes factor functions as an Occam's razor, the (linear) null model receives most evidence.

Even though we found evidence for a linear effect, there is still posterior model uncertainty and therefore we computed the Bayes factors between the one-sided models \eqref{onesided} under the nonlinear model $M_1$. This resulted in Bayes factors for the consistently positive, consistently negative, and the complement model against the unconstrained model of $B_{(1,\text{pos})u}=\frac{.825}{.140}=5.894$, $B_{(2,\text{pos})u}=\frac{.000}{.140}=0.000$, and $B_{(1,\text{comp})u}=\frac{.175}{.720}=0.242$, and thus most evidence for a consistently positive effect $B_{(1,\text{pos})(1,\text{neg})}\approx\infty$ and $B_{(1,\text{pos})(1,\text{neg})}\approx24.28$. These results are confirmed when checking the slopes of the posterior draws of the nonlinear mean function in Figure \ref{appfigure} (upper left panel).

%Fig. 2 Relation between the number of Facebook friends and the
%normalized gray matter (GM) density at the peak coordinate of the right entorhinal cortex. A positive correlation indicates that people with many Facebook friends have denser gray matter in the right entorhinal cortex. Data are replotted from Kanai, Bahrami, Roylance, and Rees (in press)

\begin{table}[t]
\begin{center}
\caption{Log Bayes factors for the linear model versus the nonlinear model using different prior scales $s_{\xi}$.}
\begin{tabular}{lccccccccccc}
\hline 
%& & \multicolumn{3}{c}{\mbox{$\log(B_{01})$}}\\
& sample size & $s_{\xi}=\mbox{e}^{-2}$ & $s_{\xi}=\mbox{e}^{-1}$ & $s_{\xi}=1$\\
\hline
Fb friends \& grey matter & 41 & 0.508 & 0.917 & 1.45\\
Age \& knowing gay & 63 & -37.7 & -38.3 & -38.1\\
Past activity \& waiting time & 500 & -0.776 & -0.361 & 0.394\\
Mother's IQ \& child test scores & 434 & -2.46 & -2.07 & -1.38\\
\hline
\end{tabular}\label{appresults}
\end{center}
\end{table}

\subsection{Sociology: Age and attitude towards gay}
We consider data presented in \cite{Gelman:2014} from the 2004 National Annenberg Election Survey containing respondents' age, sex, race, and attitude on three gay-related questions from the 2004 National Annenberg Election Survey. Here we are interested in the relationship between age and the proportion of people who know someone who's gay ($n=63$). It may be expected that older people may know less people who are gay and thus a negative relationship may be expected. Here we test whether the relationship between these variables is linear or not. In the case of a nonlinear relationship we also perform the one-sided test whether the relationship is consistently positive, negative, or neither. Again the employed model also has an intercept.

When setting the prior scale to a medium deviation from linearity, the logarithm of the Bayes factor between the linear model against the nonlinear model was approximately equal to $-38.3$, which corresponds to a Bayes factor of 0. This implies convincing evidence for a nonlinear effect. When using a small or large prior scale, the Bayes factors result in the same conclusion (Table \ref{appresults}). Figure \ref{appfigure} (upper right panel) displays the data (black circles) and 50 posterior draws of the mean function, which have clear nonlinear curves which fit the observed data.

Next we computed the Bayes factors for the one-sided test which results in decisive evidence for the complement model that the relationship is neither consistently positive nor consistently negative, with $B_{(1,\text{comp})(1,\text{pos})}=\infty$ and $B_{(1,\text{comp})(1,\text{neg})}=\infty$. This is confirmed when checking the posterior draws of the mean function in Figure \ref{appfigure} (upper right panel). We see a slight increase of the proportion of respondents who know someone who's gay towards the age of 45, and a decrease afterwards.

\subsection{Social networks: inertia and dyadic waiting times}
In dynamic social network data it is often assumed that actors in a network have a tendency to continue initiate social interactions with each other as a function of the volume of past interactions. This is also called inertia. In the application of the relational event model \cite{Butts:2008}, it is often assumed that the expected value of the logarithm of the waiting time between social interactions depends linearly on the number of past social interactions between actors. Here we consider relational (email) data from the Enron e-mail corpus \citep{Cohen:2009}. We consider a subset of the last $n=500$ emails (excluding 4 outliers) in a network of 156 employees in the Enron data \citep{Cohen:2009}. We use a model with an intercept.

Based on a medium prior scale under the nonlinear model, the logarithm of the Bayes factor between the linear model against the nonlinear model equals $\log(B_{01})=-0.361$, which corresponds to $B_{10}=1.43$, implying approximately equal evidence for both models. The posterior probabilities for the two models would be 0.412 and 0.588 for model $M_0$ and $M_1$, respectively. When using the small and large prior scale the Bayes factors are similar (Table \ref{appresults}), where the direction of the evidence flips towards the null when using a large prior scale. This could be interpreted as that a large deviation from linearity is least likely. Based on the posterior draws of the mean function in Figure \ref{appfigure} (lower left panel) we also see an approximate linear relationship. The nonlinearity seems to be mainly caused by the larger observations of the predictor variable. As there are relatively few large observations, the evidence is inconclusive about the nature of the relationship (linear or nonlinear). This suggests that more data would be needed in the larger region of the predictor variable. 

The Bayes factors for the one-sided tests yield most evidence for a consistent decrease but the evidence is not conclusive in comparison to the complement model with $B_{(1,\text{neg})(1,\text{pos})}=\infty$ and $B_{(1,\text{neg})(1,\text{comp})}=5.14$. This suggests that there is most evidence that dyads (i.e., pairs of actors) that have been more active in the past will communicate more frequently.

\subsection{Education: Mother's IQ and child test scores}
In \cite{GelmanHill} the relationship between the mother's IQ and the test scores of her child is explored while controlling for whether the mother finished her high school. The expectation is that there is a positive relationship between the two key variables, and additionally there may be a positive effect of whether the mother went to high school. Here we explore whether the relationship between the mother's IQ and child test scores is linear. An ANCOVA model is considered with an intercept and a covariate that is either 1 or 0 depending on whether the mother finished high school or not\footnote{As discussed by \cite{GelmanHill} an interaction effect could also be reasonable to consider. Here we did not add the interaction effect for illustrative purposes. We come back to this in the Discussion.}.

Based on a medium prior scale, we obtain a logarithm of the Bayes factor for $M_0$ against $M_1$ of $-2.07$. This corresponds to a Bayes factor of $B_{10}=7.92$ which implies positive evidence for the nonlinear model. Table \ref{appresults} shows that the evidence for $M_1$ is slightly higher (lower) when using a smaller (larger) prior scale. This suggests that a small deviation from linearity is more likely than a large deviation a posteriori.

Next we computed the Bayes factors for testing whether the relationships are consistently increasing, consistently decreasing, or neither. We found clear evidence for a consistently increasing effect with Bayes factors equal to $B_{(1,\text{pos})(1,\text{neg})}=\infty$ and $B_{(1,\text{pos})(1,\text{comp})}=13.3$. This implies that, within this range of the predictor variable, a higher IQ of the mother always results in a higher expected test score of the child. This is also confirmed from the random posterior curves in Figure \ref{appfigure} (lower right panel) where we correct for whether the mother finished high
school (blue lines) or not (green lines).

\section{Discussion}
In order to make inferences about the nature of the relationship between two variables principled statistical tests are needed. In this paper a Bayes factor was proposed that allows one to quantify the relative evidence in the data between a linear relationship and a nonlinear relationship, possibly while controlling for certain covariates. The test is useful (i) when one is interested in assessing whether the relationship between variables is more likely to be linear or more likely to be nonlinear, and (ii) to determine whether a certain relationship is linear after transformation.

A Gaussian process prior with a square exponential kernel was used to model the nonlinear relationship under the alternative (nonlinear) model. The model was parameterized similar as Zellner's $g$ prior to make inferences that are invariant of the scale of the dependent variable and predictor variable. Moreover the Gaussian process was parameterized using the reciprocal of the scale length parameter which controls the smoothness of the nonlinear trend so that the linear model would be obtained when setting this parameter equal to 0. Moreover a standardized scale for this parameter was proposed to quantify the deviation from linearity under the alternative model.

In the case of a nonlinear effect a Bayes factor was proposed for testing whether the effect was consistently positive, consistently negative, or neither. This test can be seen as a nonlinear extension of Bayesian one-sided testing. Unlike the linear one-sided test, the Bayes factor depends on the prior scale for the nonlinear one-sided test. Thus, the prior scale also needs to be carefully chosen for the one-sided test depending on the expected deviation from linearity.

As a next step it would be useful to extend the methodology to correct covariates that have a nonlinear effect on the outcome variable \citep[e.g., using additive Gaussian processes;][]{Cheng:2019,Duvenaud:2011}, to test nonlinear interaction effects, or to allow other kernels to model other nonlinear forms. We leave this for future work.

%\section*{Acknowledgements}
%The author was supported by an ERC Starting Grant (758791). The author is grateful to Andrew Gelman for sharing his data.

\bibliographystyle{apacite}
\bibliography{refs_mulder}

\end{document}